\documentclass[%
reprint,
showpacs,preprintnumbers,
nofootinbib,
 amsmath,amssymb,
 aps,
]{revtex4-1}

\usepackage{graphicx}% Include figure files
\usepackage{dcolumn}% Align table columns on decimal point
\usepackage{bm}% bold math
\usepackage[normalem]{ulem}  % \sout{old text} for strikeout
\usepackage[dvips]{color} % For blue in-text comments and additions

\renewcommand\sout{\bgroup \color{red} \ULdepth=-.5ex \ULset}
\newcommand{\im}{\text{Im }}

\newcommand{\bra}[1]{\langle \, #1 \, |}
\newcommand{\ket}[1]{| \, #1 \, \rangle}

\newcommand{\PDG}{Nakamura:2010zzi}

\begin{document}

%\preprint{APS/123-QED}

\title{Determination of the $\pi\Sigma$ scattering lengths from the weak decays of $\Lambda_c$}% Force line breaks with \\
%\thanks{A footnote to the article title}%

\author{Tetsuo Hyodo}
 \email{hyodo@th.phys.titech.ac.jp}
\affiliation{%
Department of Physics, Tokyo Institute of Technology, 
Meguro, Tokyo 152-8551, Japan
}%

\author{Makoto Oka}%
\affiliation{%
Department of Physics, Tokyo Institute of Technology, 
Meguro, Tokyo 152-8551, Japan
}%

%\collaboration{MUSO Collaboration}%\noaffiliation

\date{\today}% It is always \today, today,
             %  but any date may be explicitly specified

\begin{abstract}
The scattering lengths of the $\pi\Sigma$ systems are key quantities in understanding the structure of the $\Lambda(1405)$ resonance and the subthreshold extrapolation of the $\bar{K}N$ interaction. We demonstrate that the $\pi\Sigma$ scattering lengths can be extracted from the threshold cusp phenomena in the weak $\Lambda_c\to \pi\pi\Sigma$ decays, analogously with Cabibbo's method for determination of the $\pi\pi$ scattering length. We show that the substantial cusp effect should be observed in the spectrum, when the $\pi\Sigma$ interaction in $I=0$ is strongly attractive to generate a near-threshold singularity, such as a bound state or a virtual state.

\end{abstract}

% PACS, the Physics and Astronomy
                             % Classification Scheme.
% 13.75.Gx : Pion-baryon interactions
% 13.30.Eg : Hadronic decays (Decays of baryons) 
% 14.20.Jn : Hyperons
% 13.75.Lb : Meson-meson interactions
%Use showkeys class option if keyword
                              %display desired
\maketitle

%\tableofcontents

%%%%%%%%%%%%%%%%%%%%%%%%%%%%%%%%%%%%%%%%%%%%%%%%%%%%%%%%%%%%%%%%%%%%%%%%
\section{Introduction}\label{sec:Introduction}
%%%%%%%%%%%%%%%%%%%%%%%%%%%%%%%%%%%%%%%%%%%%%%%%%%%%%%%%%%%%%%%%%%%%%%%%

% Kbar-nuclei and KbarN interaction
The possibility of kaon bound states in nuclei~\cite{PL7.288,Akaishi:2002bg} has been intensely discussed recently~\cite{Shevchenko:2006xy,Ikeda:2007nz,Shevchenko:2007zz,Yamazaki:2007cs,Dote:2008in,Dote:2008hw,Ikeda:2008ub,Ikeda:2010tk} and is one of the central subjects in forthcoming experiments at Japan Proton Accelerator Research Complex E15~\cite{Hiraiwa:2011zz}, FOPI at GSI~\cite{Suzuki:2009zze}, and AMADEUS at DA$\Phi$NE~\cite{Zmeskal:2010zz}. An important theoretical concept to describe kaonic nuclei is the two-body $\bar{K}N$-$\pi\Sigma$ interaction, which closely relates to the description of the $\Lambda(1405)$ resonance in the $I=0$ amplitude. Theoretical models of the $\bar{K}N$-$\pi\Sigma$ scattering amplitudes have been constrained by the total cross sections of low-energy $K^-p$ scattering into various final states and the threshold branching ratios of the $K^-p$ channel. Recently, due to intensive experimental activities, new data of the $\pi\Sigma$ invariant mass distributions are becoming available, and the analysis of the energy level and width of the kaonic hydrogen provides precise information regarding the $K^-p$ scattering length~\cite{Bazzi:2011zj}.

% subthreshold extrapolation and Lambda(1405)
In this way, there are many experimental data around the $\bar{K}N$ threshold, while the amplitude at far below the threshold is not well constrained. This means that the $\bar{K}N$-$\pi\Sigma$ interaction relevant to the study of the $\bar{K}$-nucleus systems with strong binding is achieved only through the subthreshold extrapolation from the $\bar{K}N$ threshold. For instance, the chiral unitary model~\cite{Kaiser:1995eg,Oset:1998it,Oller:2000fj,Lutz:2001yb,Hyodo:2011ur}, a coupled-channels approach with the low-energy interaction constrained by chiral symmetry, has revealed an interesting two-pole structure~\cite{Jido:2003cb}. This indicates that the nominal $\Lambda(1405)$ resonance is not a single state but a superposition of two states with different properties. It is shown that this two-pole structure has substantial influence for the $\bar{K}N$ interaction~\cite{Hyodo:2007jq} and the $\bar{K}NN$-$\pi\Sigma N$ system~\cite{Ikeda:2010tk}. However, it is found that the position of the lower energy pole is sensitive to the details of the model, due to the lack of the information of the $\pi\Sigma$ interaction~\cite{Borasoy:2004kk,Borasoy:2005ie,Oller:2005ig,Oller:2006jw,Borasoy:2006sr,Ikeda:2011dx}. 

% pi Sigma scattering length
The scattering length characterizes the property of the two-body interaction. The attractive/repulsive nature of the interaction is reflected in the sign of the scattering length. The magnitude is sensitive to the singularity of the amplitude close to the threshold. If there is a bound (virtual) state, the scattering length becomes large and negative (positive), as is known for the ${}^3S_1$ (${}^1S_0$) $NN$ scattering~\cite{BlattWeisskopf}. In the case of the $\bar{K}N$-$\pi\Sigma$ interaction, it is shown that the position of the lower energy pole of $\Lambda(1405)$ is closely related to the $\pi\Sigma$ scattering length with isospin $I=0$~\cite{Ikeda:2011dx}. Therefore, it is highly desirable to determine the $\pi\Sigma(I=0)$ scattering length to impose further constraints on the low-energy $\bar{K}N$-$\pi\Sigma$ interaction. There are several ways to extract the hadron scattering length in experiments. The standard method is to measure the energy shift and the width of the hadronic atom state~\cite{Beer:2005qi,Adeva:2005pg,Gasser:2007zt,Bazzi:2011zj} or to extrapolate the low-energy phase-shift data down to the threshold~\cite{Batley:2007zz}. Unfortunately, both methods cannot be applied to the $\pi\Sigma$ case.

% Cabibbo's method 
Recently, an alternative method to determine hadron scattering lengths was proposed by Cabibbo~\cite{Cabibbo:2004gq,Cabibbo:2005ez}, in which the cusp effect at the $\pi^+\pi^-$ threshold is used to extract the $\pi\pi$ scattering length in the $K^+\to \pi^+\pi^0\pi^0$ decay. The structure around the cusp at the $\pi^{+}\pi^{-}$ threshold is related to the scattering amplitude of the $\pi^{+}\pi^{-}\to\pi^{0}\pi^{0}$ process, which is nothing but the $\pi\pi$ scattering length (the relationship between the scattering length and the cusp phenomena in the $\pi\pi$ spectrum has been discussed also in Ref.~\cite{Meissner:1997fa}). The cusp effect is indeed observed in the high-statistics experimental data~\cite{Batley:2005ax,Batley:2010fj} which leads to the determination of the $\pi\pi$ scattering length.

% proposal for pi Sigma length in Lambda_c decay
In this paper, we propose to apply a similar methodology to the $\pi\Sigma$ scattering lengths by use of the $\Lambda_c\to \pi\pi\Sigma$ decays. This decay process has been measured as the $\Lambda_{c}$ peak in the three-body invariant mass spectrum in Refs.~\cite{Barlag:1992jz,Kubota:1993pw,Avery:1993ri,Frabetti:1994kt,VazquezJauregui:2008eg}. The branching ratios to the relevant decay channels are reported as a few percents by the Particle Data Group (PDG)~\cite{\PDG}, so the extraction would be feasible at Belle and Babar where a huge amount of $\Lambda_c$ is produced. 

% paper organization
The contents of this paper are as follows. In Sec.~\ref{sec:analysis}, we discuss the general property of the threshold cusp phenomena in the $\Lambda_{c}\to \pi\pi\Sigma$ process. We introduce an expansion scheme to extract the scattering length, keeping the value of the scattering length unspecified. In Sec.~\ref{sec:estimation}, we present theoretical estimations of the decay process and show the expected spectra for several values of the scattering lengths. The last section is devoted to the summary.

%%%%%%%%%%%%%%%%%%%%%%%%%%%%%%%%%%%%%%%%%%%%%%%%%%%%%%%%%%%%%%%%%%
\section{Analysis of the cusp phenomena}\label{sec:analysis}
%%%%%%%%%%%%%%%%%%%%%%%%%%%%%%%%%%%%%%%%%%%%%%%%%%%%%%%%%%%%%%%%%%

Here we analyze the threshold cusp phenomena in the $\Lambda_c\to \pi\pi\Sigma$ decay. We review the general argument of the determination of the scattering length from the threshold cusp effect and summarize the possible $\pi\Sigma$ channels in the $\Lambda_c$ decay where we have a chance to observe the cusp structure. We then discuss the method to extract the scattering length with minimal model assumptions by expanding the amplitude in terms of the momentum variable. We also present an extension of the framework to the case with complex amplitudes.

%------------------------------
\subsection{$\Lambda_c\to \pi\pi\Sigma$ decay and the $\pi\Sigma$ scattering length}\label{subsec:idea}

Let us consider the weak decay of $\Lambda_c$ into two pions and one $\Sigma$ baryon. We denote a $\pi\Sigma$ pair in the final state as $(\pi\Sigma)_l$ and concentrate on the threshold energy region of this pair. In this case, the additional primary pion has a large momentum because of the large phase space of the decay process, as we see in Appendix~\ref{sec:kinematics}. Choosing an appropriate charged state for the $(\pi\Sigma)_l$ pair, other charge combinations of the $\pi\Sigma$ state may be allowed at slightly higher energies than the $(\pi\Sigma)_l$ threshold due to the isospin violation in the particle masses. We denote the higher energy channel of the $\pi\Sigma$ pair as $(\pi\Sigma)_h$. The specific charge states of $(\pi\Sigma)_l$ and $(\pi\Sigma)_h$ will be given in Sec.~\ref{subsec:threshold}, in connection with possible experimental observations.

%--figure---------------------------------
\begin{figure}[bp]
    \centering
    \includegraphics[width=0.45\textwidth,clip]{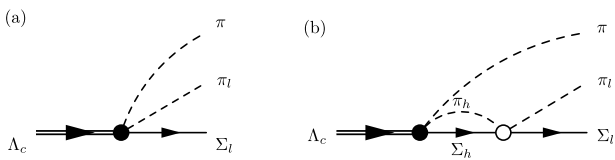}
    \caption{\label{fig:diagrams}
    Decay diagrams for the $\Lambda_c\to \pi(\pi\Sigma)_l$ process: 
    (a) direct decay and
    (b) decay through the $(\pi\Sigma)_h$ state.
    Solid circles denote the weak process of $\Lambda_c\to \pi\pi \Sigma$
    and the open circle represents the scattering amplitude of the 
    $(\pi\Sigma)_h\to (\pi\Sigma)_l$ process.}
\end{figure}%
%--figure---------------------------------

The dominant part of the $\Lambda_c$ decay is given by the direct process:
\begin{align}
    \Lambda_c
    \to & \pi(\pi\Sigma)_l ,
    \nonumber
\end{align}
as shown in Fig.~\ref{fig:diagrams}(a). In addition, we may have the final-state interaction term with the intermediate $(\pi\Sigma)_h$ state as
\begin{align}
    \Lambda_c
    \to & \pi(\pi\Sigma)_h
    \to \pi(\pi\Sigma)_l,
    \nonumber
\end{align}
which is depicted in Fig.~\ref{fig:diagrams}(b). Because of the mass difference, the $(\pi\Sigma)_h$ threshold appears in the mass spectrum of the $(\pi\Sigma)_l$ channel slightly above the $(\pi\Sigma)_l$ threshold. The cusp phenomenon occurs at this higher energy threshold.

It is important to note that the amplitude in Fig.~\ref{fig:diagrams}(b) contains the vertex of $(\pi\Sigma)_h\to (\pi\Sigma)_l$. The threshold cusp takes place with the vanishing momentum of the $(\pi\Sigma)_h$ state, and it contains the information of the on-shell scattering amplitude of the $(\pi\Sigma)_h\to (\pi\Sigma)_l$ process. Following Ref.~\cite{Cabibbo:2005ez}, we define the off-diagonal $\pi\Sigma$ scattering length as (the real part of) the amplitude $f_{h\to l}$ at the higher energy threshold $W=W_{\text{th}}$:
\begin{align}
    a_{h\to l}
    \equiv &
    f_{h\to l}(W=W_{\text{th}})
    \label{eq:def} ,
\end{align}
which is reflected in the behavior of the cusp structure. In the following, we show how to extract the scattering length~\eqref{eq:def} along the same line with Ref.~\cite{Cabibbo:2004gq}.

%------------------------------
\subsection{Possible decay modes for the scattering length}\label{subsec:threshold}

%--figure---------------------------------
\begin{figure}[bp]
    \centering
    \includegraphics[width=0.4\textwidth,clip]{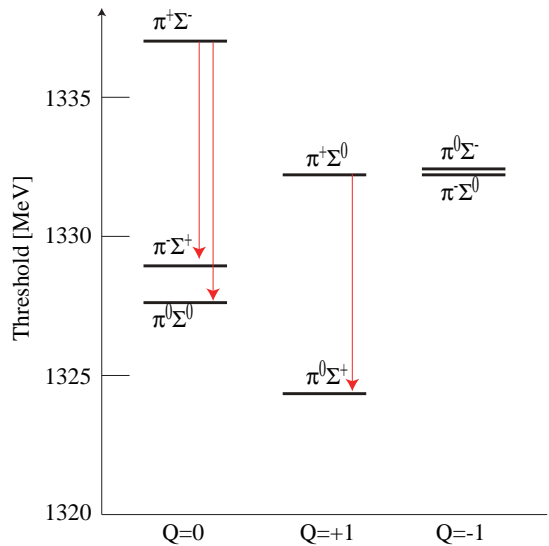}
    \caption{\label{fig:pSthreshold}
    (Color online)
    Threshold energies of the $\pi\Sigma$ states with charge $Q$. The arrows indicate the possible decay modes in which visible cusp structure may appear.}
\end{figure}%
%--figure---------------------------------

We first consider the energy levels of the $\pi\Sigma$ channels with different charge combinations to specify the $(\pi\Sigma)_l$ and $(\pi\Sigma)_h$ channels. Since the charge $Q=\pm 2$ states have no coupled channels, they are not relevant in the present discussion. The threshold energy levels for the $Q=0$, $+1$, and $-1$ channels are shown in Fig.~\ref{fig:pSthreshold}. Unlike the $\pi\pi$ case where the masses of $\pi^+$ and $\pi^-$ are identical, $\Sigma^-$ is heavier than $\Sigma^+$, so the $\pi\Sigma$ channels show a rich spectrum. Among them, there are about 10-MeV mass differences in the following transitions:
\begin{align}
    \pi^{+}\Sigma^-
    \to &\pi^{-}\Sigma^+ ,
    \quad \pi^{+}\Sigma^-
    \to &\pi^{0}\Sigma^0 ,
    \quad \pi^{+}\Sigma^0
    \to &\pi^{0}\Sigma^+ .\nonumber
\end{align}
We expect to observe the cusp effects by regarding these decays as $(\pi\Sigma)_h\to (\pi\Sigma)_l$. 

Using the isospin decomposition with the phase convention given in Eq.~\eqref{eq:phaseconvention}, we can express the scattering lengths in these channels as
\begin{align}
    a^{-+}
    = & \tfrac{1}{3}a^{0}
    -\tfrac{1}{2}a^{1}
    +\tfrac{1}{6}a^{2}+\dotsb  ,
    \nonumber \\*
    a^{00}
    =&\tfrac{1}{3}a^{0}
    -\tfrac{1}{3}a^{2}+\dotsb  ,
    \nonumber \\
    a^{0+}
    =&-\tfrac{1}{2}a^{1}
    +\tfrac{1}{2}a^{2}+\dotsb ,
    \nonumber 
\end{align}
where the scattering lengths are labeled by the charges of the final states, $a^{I}$ is the scattering length with isospin $I$ in the isospin basis, and the ellipses represent the isospin breaking corrections.\footnote{In principle, $a^{1}$ can be complex, since the $\pi\Lambda$ channel is open at the $\pi\Sigma$ threshold. However, the transition $\pi\Sigma(I=1)\to \pi\Lambda$ vanishes at the leading order in chiral perturbation theory. In the following we assume the imaginary part is negligible.} Although we have three equations, they are not linearly independent if we neglect the isospin breaking corrections as
\begin{align}
    a^{-+} - a^{00} 
    = &a^{0+} +\dotsb .
    \nonumber 
\end{align}
Therefore, even if we extract $a^{-+}$, $a^{00}$, and $a^{0+}$  in experiments, in order to determine all three $a^{I}$s, we need one additional input. We may adopt chiral perturbation theory (ChPT) or the lattice QCD simulation for the determination of $a^{I=2}$ where the interaction is presumably weak and repulsive.

Let us check the decay modes of $\Lambda_c$ which lead to these rescattering amplitudes. Since the charge of $\Lambda_c$ is $Q=+1$, the possible decay modes are 
\begin{align}
    a^{-+} :& \Lambda_c\to \pi^+(\pi^+\Sigma^-) 
    \to \pi^+(\pi^-\Sigma^+) ,
    \label{eq:amp} \\
    a^{00} :& \Lambda_c\to \pi^+(\pi^+\Sigma^-) 
    \to \pi^+(\pi^0\Sigma^0) ,
    \label{eq:a00} \\
    a^{0+} :& \Lambda_c\to \pi^0(\pi^+\Sigma^0) 
    \to \pi^0(\pi^0\Sigma^+) .
    \label{eq:a0p}
\end{align}
For later convenience, we refer to these modes as the extracted scattering lengths, i.e., $\Lambda_c\to \pi^+(\pi^+\Sigma^-) \to \pi^+(\pi^-\Sigma^+)$ is called mode $a^{-+}$. The experimental data~\cite{\PDG} of the branching ratios $\Gamma_i/\Gamma$ for these channels are summarized in Table~\ref{tbl:ratios}. 

\begin{table}[tb]
\caption{\label{tbl:ratios} Decay branching ratios $\Gamma_i/\Gamma$ from PDG~\cite{\PDG}.}
\begin{ruledtabular}
\begin{tabular}{lcc}
mode & $\Lambda_c\to \pi(\pi\Sigma)_h$ & $\Lambda_c\to \pi(\pi\Sigma)_l$ \\
\colrule
$a^{-+}$ & $1.7\pm 0.5$ \% & $3.6\pm 1.0$ \% \\
$a^{00}$ & $1.7\pm 0.5$ \% & $1.8\pm 0.8$ \% \\
$a^{0+}$ & $1.8\pm 0.8$ \% & Not known \\
\end{tabular}
\end{ruledtabular}
\end{table}%

The mass spectrum is extracted from the $\Lambda_c\to \pi(\pi\Sigma)_l$ mode, while the $\Lambda_c\to \pi(\pi\Sigma)_h$ mode is used to normalize the strength of the amplitude, as we will see below. Fortunately, most decay modes are experimentally observed, with the branching ratio of the order of several percents. If the precise measurement of the $\pi\Sigma$ spectrum around the threshold is performed, it will be feasible to extract the scattering lengths by the method explained in Sec.~\ref{subsec:expansion}.

%------------------------------
\subsection{Threshold cusp effect}\label{subsec:cusp}

Let us consider the $\Lambda_{c}$ decay process in Fig.~\ref{fig:diagrams}, paying attention to the imaginary part of the loop function. To concentrate on the cusp phenomena, in this section, we simply ignore the diagrams other than those in Fig.~\ref{fig:diagrams}, assuming that the other final-state interactions  are slowly varying in the relevant energy region. The decay width of the process $\Lambda_{c}\to \pi(\pi\Sigma)_{l}$ can be calculated as 
\begin{align}
    \Gamma
    = &
    \int d\Pi_3 \ \Sigma|\mathcal{M}|^2 ,
    \nonumber
\end{align}
where $\mathcal{M}$ is the relativistic scattering amplitude of the process, $d\Pi_3$ is the three-body phase space and $\Sigma$ denotes the spin summation. We denote the mass of the $\Sigma_h$ ($\pi_h$) by $M_h$ ($m_h$), and the invariant mass of the $(\pi\Sigma)_l$ system by $W$. The mass spectrum of the decay with respect to $W$ is given by
\begin{align}
    \frac{d\Gamma}{dW}
    = &
    \int d\tilde{\Pi} \ \Sigma |\mathcal{M}|^2,
    \nonumber
\end{align}
with $d\tilde{\Pi}=d\Pi_3/dW$. Since we are interested in the cusp structure which arises from the nonanalytic behavior of $W$, we concentrate on the amplitude as a function of $W$, assuming that the dependence on the other kinematical variables can be factorized. Integrating the other variables, we arrive at the expression
\begin{align}
    \frac{d\Gamma}{dW}
    =& \frac{M_{l} W}{16\pi^3M_{\Lambda_c}}
    \Sigma|\mathcal{M}|^2
    \int d\omega_{l} 
    \Theta(1-A^2)
    \label{eq:spectrum} ,
    \\
    A=& 
    \frac{(M_{\Lambda_c}-\omega_{\pi}
    -\omega_{l})^2
    -M_{l}^2-p_{\pi}^2-p_{l}^2}{2p_{\pi}p_{l}} 
    \nonumber , \\
    \omega_{\pi}
    = &
    \frac{M_{\Lambda_c}^2-W^2+m_{\pi}^2}{2M_{\Lambda_c}}
    \nonumber , \\
    p_{\pi}
    = & 
    \sqrt{\omega_{\pi}^2-m_{\pi}^2} ,
    \quad 
    p_{l}
    = 
    \sqrt{\omega_{l}^2-m_{l}^2}
    \nonumber ,
\end{align}
where $M_{\Lambda_{c}}$ is the mass of the $\Lambda_{c}$, $\omega_{l}$ ($\omega_{\pi}$) and $p_{l}$ ($p_{\pi}$) are the energy and momentum of $\pi_{l}$ (the primary pion), and $\Theta(x)$ is the step function.

To appreciate the threshold cusp effect, we focus on the property of the loop function $G(W)$ in Fig.~\ref{fig:diagrams}(b). The loop function can be written in the spectral representation as
\begin{align}
    G(W)
    = &
    \frac{1}{2\pi}\int_{W_{\text{th}}}^{\infty}dW^{\prime}
    \frac{\rho(W^{\prime})}{W-W^{\prime}+i\epsilon}
    +\text{(subtractions)} ,
    \nonumber 
\end{align}
where the threshold energy is $W_{\text{th}}= M_h+m_h$ and the phase-space factor $\rho(W)$ is given by
\begin{align}
    \rho(W)
    = &
    2M_h \frac{q(W)}{4\pi W} ,
    \nonumber
\end{align}
with the three-momentum function
\begin{align}
    q(W) 
    =& \frac{\sqrt{[W^2-(M_h-m_h)^2](W^2-W_{\text{th}}^2)}}{2W} 
    \nonumber .
\end{align}
The real part of the loop function $G(W)$ depends on the subtractions, but the imaginary part can be determined only by the kinematics. Since the scattering amplitude with the diagram in Fig.~\ref{fig:diagrams} contains the loop function, we obtain the condition for the amplitude
\begin{align}
    \im \mathcal{M}(W)
    \propto \im G(W)
    = &
    -\frac{M_hq(W)}{4\pi W}
    \Theta(W-W_{\text{th}})
    \label{eq:imaginarypart} .
\end{align}
This means that the imaginary part of the amplitude appears suddenly at the threshold. 

Considering the two diagrams shown in Fig.~\ref{fig:diagrams}, we can decompose the $\Lambda_c\to \pi(\pi\Sigma)_l$ amplitude into two parts:
\begin{align}
    \mathcal{M}(W)
    = &
    \mathcal{M}_0(W) + i\tilde{\mathcal{M}}_1(W) q(W)
    \quad  \text{for}\quad W>W_{\text{th}}
    \label{eq:amplitude} ,
\end{align}
with $\mathcal{M}_0(W)$ and $\tilde{\mathcal{M}}_1(W)$ being analytic functions of $W$ at the threshold $W_{\text{th}}$. Here we also assume that $\mathcal{M}_0(W)$ and $\tilde{\mathcal{M}}_1(W)$ have no imaginary part, and the extension to the complex amplitudes is discussed in Sec.~\ref{subsec:extension}. The amplitude~\eqref{eq:amplitude} satisfies the condition~\eqref{eq:imaginarypart}, because the $(\pi\Sigma)_h$ loop is the only source of the imaginary part in this process. Physically, the amplitude of the direct process (a) and the real part of the indirect process (b) are included in $\mathcal{M}_0$, and the amplitude corresponding to the imaginary part of the indirect process (b) is included in $i\tilde{\mathcal{M}}_1q(W)$. The function $q(W)$ can be analytically continued to $W<W_{\text{th}}$ where it becomes pure imaginary. For later convenience, we define the dimensionless quantity
\begin{align}
    \delta 
    \equiv &
    \frac{\sqrt{[W^2-(M_h-m_h)^2](W_{\text{th}}^2-W^2)}}{2m_h W}
    =\frac{iq(W)}{m_h}
    \label{eq:delta} ,
\end{align}
so $\delta$ is real (imaginary) below (above) the threshold and $\delta^{2}=-|\delta|^{2}$ for $W>W_{\text{th}}$.\footnote{There is another branch point at $W= M_h-m_h$, but we will not consider this because the energy region cannot be reached in our analysis of $\pi\Sigma$ spectrum around the threshold.} The amplitude then can be rewritten as
\begin{align}
    \mathcal{M}(W)
    = &
    \mathcal{M}_0(W) + \tilde{\mathcal{M}}_1(W) m_h\delta
    \label{eq:amplitude2} .
\end{align}
Because of the property of $\delta$, the second term of Eq.~\eqref{eq:amplitude2} interferes with the first term below the threshold, while such an interference does not occur above the threshold. 

Using the expression~\eqref{eq:amplitude2} and $\mathcal{M}^{*}=\mathcal{M}_0 + \tilde{\mathcal{M}}_1 m_h\delta^{*}$, we calculate the amplitude square as
\begin{widetext}
\begin{align}
    |\mathcal{M}|^2 
    = &
    \begin{cases}
      (\mathcal{M}_0)^2 + (\tilde{\mathcal{M}}_1m_h)^2|\delta|^2 
      & \text{for}\quad W>W_{\text{th}} \\
      (\mathcal{M}_0)^2 
      +2\mathcal{M}_0 \tilde{\mathcal{M}}_1 m_h\delta
      + (\tilde{\mathcal{M}}_1 m_h)^2\delta^2 
      & \text{for}\quad W< W_{\text{th}}
    \end{cases}
    \label{eq:ampsquare} .
\end{align}
\end{widetext}
Note that $\delta$ vanishes at the threshold, so the spectrum is continuous at $W=W_{\text{th}}$. On the other hand, its derivative with respect to $W$ is not continuous:
\begin{align}
    & \left.\frac{d|\mathcal{M}|^2}{dW}
    \right|_{W\to W_{\text{th}}-0}
    -\left.\frac{d|\mathcal{M}|^2}{dW}
    \right|_{W\to W_{\text{th}}+0} \nonumber \\
    \propto &\
    -\frac{2\mathcal{M}_0 
    \tilde{\mathcal{M}}_1m_h M_h}{M_h+m_h}\frac{1}{\delta}
    +\mathcal{O}(\delta)
    \nonumber .
\end{align}
This means that the $(\pi\Sigma)_l$ spectrum is continuous but not smooth at the $(\pi\Sigma)_h$ threshold. This singular behavior is called the threshold cusp. Although we have assumed that $\mathcal{M}_0$ is real in the present case, the mechanism of the threshold cusp is the same even if $\mathcal{M}_0$ has an imaginary part, since the discontinuity is caused by the nonanalytic term $\delta$ as a function of $W$, as shown in Sec.~\ref{subsec:extension}. Thus, the cusp always occurs at the threshold due to the kinematical condition, but the simple interference pattern in Eq.~\eqref{eq:ampsquare} is realized only for the real valued $\mathcal{M}_0$ amplitude.

%------------------------------
\subsection{Expansion of the amplitude and spectrum}
\label{subsec:expansion}

In order to extract the scattering length, it is desirable to parametrize the amplitudes around the threshold $W_{\text{th}}$. The aim of this section is to relate the $\pi\Sigma$ scattering length to the near-threshold behavior of the spectrum. Since we concentrate on the energy region around the $(\pi\Sigma)_h$ threshold, the variable $\delta$ should be small. As we will see below, the relevant energy region is $|W-W_{\text{th}}|\lesssim 10$ MeV, where the value of $\delta$ is $|\delta|\lesssim 0.37$. This means that the $\mathcal{O}(\delta^3)$ error is about 5 \% level.

Since the $\mathcal{M}_0$ amplitude is an analytic function of $W$, it has only even powers in the $\delta$ expansion: 
\begin{align}
    \mathcal{M}_0(\delta)
    = &
    \mathcal{M}_0^{(0)}
    +\mathcal{M}_0^{(2)} \delta^2
    +\mathcal{O}(\delta^4) .
    \label{eq:M0expand}
\end{align}
The $\tilde{\mathcal{M}}_1$ part of the amplitude is related to the scattering length. Identifying this contribution as the imaginary part of the diagram in Fig.~\ref{fig:diagrams}(b), we can write the amplitude above the threshold as 
\begin{align}
    \tilde{\mathcal{M}}_1(W) q
    = &
    -\mathcal{M}_0^h(W)
    \frac{M_hq}{4\pi W}
    \mathcal{M}_{h\to l}(W) 
    \nonumber \\
    \tilde{\mathcal{M}}_1(W) 
    = &
    \mathcal{M}_0^h(W)
    f_{h\to l}(W)
    \nonumber ,
\end{align}
where $\mathcal{M}_0^h$ is the amplitude of the weak $\Lambda_c\to \pi(\pi\Sigma)_h$ decay. The amplitude $\mathcal{M}_{h\to l}$ represents the transition amplitude for the $(\pi\Sigma)_h\to(\pi\Sigma)_l$ process and $f_{h\to l}$ is its nonrelativistic counterpart. Note that all the final-state interactions \textit{via} the strong interaction are implicitly included in $f_{h\to l}$. The $\mathcal{M}_0^h$ amplitude is an analytic function of $W$ so 
\begin{align}
    \mathcal{M}_0^h(\delta)
    = &
    \mathcal{M}_0^{h(0)}
    +\mathcal{M}_0^{h(2)} \delta^2
    +\mathcal{O}(\delta^4) .
    \label{eq:Mhexpand}
\end{align}
The low-energy behavior of the nonrelativistic scattering amplitude can be parametrized by the scattering length $a_{h\to l}$. According to Eq.~\eqref{eq:def}, the scattering length is defined at the higher energy threshold, so we expand it in terms of $\delta$ as
\begin{align}
    f_{h\to l}(q)
    = &
    \left(\frac{1}{a_{h\to l}}-iq +\cdots\right)^{-1} \nonumber \\
    =& a_{h\to l}+a_{h\to l}^{2}m_{h}\delta + \cdots
    \label{eq:FSIexand} .
\end{align}
Using Eqs.~\eqref{eq:Mhexpand} and \eqref{eq:FSIexand}, we can express $\tilde{\mathcal{M}}_1$ as
\begin{align}
    \tilde{\mathcal{M}}_1(\delta)
    = &
    \mathcal{M}_0^{h(0)}a_{h\to l}
    +\mathcal{M}_0^{h(0)} m_{h}a_{h\to l}^2 \delta
    +\mathcal{O}(\delta^2)
    \label{eq:M1amp} .
\end{align}
Combining Eqs.~\eqref{eq:amplitude2}, \eqref{eq:M0expand}, and \eqref{eq:M1amp}, the expansion of $\mathcal{M}$ up to $\delta^{2}$ is given by
\begin{align}
    \mathcal{M}(\delta)
    = &
    \mathcal{M}_0^{(0)}
    +
    \mathcal{M}_0^{h(0)}m_{h}a_{h\to l}\delta \nonumber \\
    &
    +(\mathcal{M}_0^{(2)}  
    +\mathcal{M}_0^{h(0)}m_{h}^{2}a_{h\to l}^2) \delta^2
    +\mathcal{O}(\delta^3)
    \nonumber .
\end{align}
The amplitude square is then given by
\begin{align}
    |\mathcal{M}|^2
    = &
    \begin{cases}
      A
      +C^{{\prime}}|\delta|^2 
      +\mathcal{O}(|\delta|^4)
      & \text{for}\quad W>W_{\text{th}} \\
      A
      +B\delta
      +C\delta^2 
      +\mathcal{O}(\delta^3)
      & \text{for}\quad W<W_{\text{th}}
    \end{cases} ,
    \label{eq:ampsquare2}
\end{align}
where the coefficients are given by 
\begin{align}
    A
    =&(\mathcal{M}_0^{(0)})^2 \nonumber , \\
    B=& 
    2\mathcal{M}_0^{(0)}\mathcal{M}_0^{h(0)}
    m_ha_{h\to l} \nonumber , \\
    C
    =&2\mathcal{M}_0^{(0)}\mathcal{M}_0^{(2)}
    +2\mathcal{M}_0^{(0)}\mathcal{M}_0^{h(0)}m_{h}^{2}a_{h\to l}^2
    \nonumber\\
    &+(\mathcal{M}_0^{h(0)})^{2}m_{h}^{2}a_{h\to l}^{2} ,
    \nonumber \\
    C^{\prime}
    =&
    -2\mathcal{M}_0^{(0)}\mathcal{M}_0^{(2)}
    -2\mathcal{M}_0^{(0)}\mathcal{M}_0^{h(0)}m_{h}^{2}a_{h\to l}^2
     \nonumber \\
    &+(\mathcal{M}_0^{h(0)})^{2}m_{h}^{2}a_{h\to l}^{2} .
    \nonumber
\end{align}
These coefficients can be extracted using the experimental spectrum, the three-body phase-space factor, and the fitting by a polynomial of $\delta$ around the threshold $W_{\text{th}}$. Since the $B$ coefficient is proportional to the scattering length, we can extract its absolute value as
\begin{align}
    |a_{h\to l}|
    = &
    \frac{|B|}{2m_h\sqrt{A}|\mathcal{M}_0^{h(0)}|}
    \nonumber ,
\end{align}
and the sign is given by 
\begin{align}
    \frac{a_{h\to l}}{|a_{h\to l}|}
    = &
    \frac{\mathcal{M}_0^{(0)}}{|\mathcal{M}_0^{(0)}|}
    \cdot \frac{\mathcal{M}_0^{h(0)}}{|\mathcal{M}_0^{h(0)}|} 
    \cdot
    \frac{B}{|B|} 
    \nonumber .
\end{align}
The $\mathcal{M}_0^{h(0)}$ coefficient can be obtained as the leading contribution for the $\Lambda_c\to \pi(\pi\Sigma)_h$ process around the threshold. It is also possible to calculate $\mathcal{M}_0^{h}$ when the information of the weak process is well under control. If we determine the relative sign of $\mathcal{M}_0^{(0)}$ and $\mathcal{M}_0^{h(0)}$, the sign of the scattering length $a_{h\to l}$ is determined by the sign of the coefficient $B$. This information, together with the fitting of the $(\pi\Sigma)_l$ spectrum in the $\Lambda_c\to \pi(\pi\Sigma)_l$ process below and above the $(\pi\Sigma)_h$ threshold by Eq.~\eqref{eq:ampsquare2}, leads to the determination of the scattering length $a_{h\to l}$. 

%------------------------------
\subsection{Extension to the complex amplitude}
\label{subsec:extension}

So far we have assumed that the weak interaction amplitudes $\mathcal{M}_{0}$ and $\tilde{\mathcal{M}}_{1}$ are real. This is valid at the leading order, but several final-state interactions may bring an imaginary part as we discuss in Appendix~\ref{sec:kinematics}. Here we consider the possible modification of the formulation by the imaginary part of these amplitudes.

Since the overall phase of the amplitude does not contribute to the spectrum, we consider a relative phase between $\mathcal{M}_{0}$ and $\tilde{\mathcal{M}}_{1}$ amplitudes. The imaginary part other than Eq.~\eqref{eq:imaginarypart} is a smooth function of $W$, so we introduce a constant phase $e^{i\theta}$ to generalize Eq.~\eqref{eq:amplitude2} as
\begin{align}
    \mathcal{M}(W)
    = &
    \mathcal{M}_0(W) + \tilde{\mathcal{M}}_1(W)e^{i\theta}
     m_h\delta
    \nonumber .
\end{align}
with $\mathcal{M}_0$ and $\tilde{\mathcal{M}}_{1}$ being real. In this case, the amplitude square is given by
\begin{widetext}
\begin{align}
    |\mathcal{M}|^2 
    = &
    \begin{cases}
      (\mathcal{M}_0)^2 
      +2\mathcal{M}_0 \tilde{\mathcal{M}}_1 m_h|\delta|
      \sin\theta
      + (\tilde{\mathcal{M}}_1m_h)^2|\delta|^2 
      & \text{for}\quad W>W_{\text{th}} \\
      (\mathcal{M}_0)^2 
      +2\mathcal{M}_0 \tilde{\mathcal{M}}_1 m_h\delta
      \cos\theta
      + (\tilde{\mathcal{M}}_1 m_h)^2\delta^2 
      & \text{for}\quad W< W_{\text{th}}
    \end{cases}
    \nonumber .
\end{align}
\end{widetext}
In this case, the linear $|\delta|$ term remains in the spectrum above threshold.

The expansion of the amplitude square in terms of $|\delta|$ is given by
\begin{align}
    |\mathcal{M}|^2
    = &
    \begin{cases}
      A
      +B^{{\prime}}|\delta|
      +C^{{\prime}}|\delta|^2 
      +\mathcal{O}(|\delta|^4)
      & \text{for}\quad W>W_{\text{th}} \\
      A
      +B\delta
      +C\delta^2 
      +\mathcal{O}(\delta^3)
      & \text{for}\quad W<W_{\text{th}}
    \end{cases} ,
    \nonumber
\end{align}
where the coefficients are given by 
\begin{align}
    A
    =&(\mathcal{M}_0^{(0)})^2 \nonumber , \\
    B=& 
    2\mathcal{M}_0^{(0)}\mathcal{M}_0^{h(0)}
    m_ha_{h\to l}\cos\theta \nonumber , \\
    B^{\prime}=& 
    2\mathcal{M}_0^{(0)}\mathcal{M}_0^{h(0)}
    m_ha_{h\to l}\sin\theta
    =B\tan \theta
     \nonumber , \\
    C
    =&2\mathcal{M}_0^{(0)}\mathcal{M}_0^{(2)}
    +2\mathcal{M}_0^{(0)}\mathcal{M}_0^{h(0)}m_{h}^{2}a_{h\to l}^2
    \cos\theta
    \nonumber\\
    &+(\mathcal{M}_0^{h(0)})^{2}m_{h}^{2}a_{h\to l}^{2}  ,
    \nonumber \\
    C^{\prime}
    =&
    -2\mathcal{M}_0^{(0)}\mathcal{M}_0^{(2)}
    -2\mathcal{M}_0^{(0)}\mathcal{M}_0^{h(0)}m_{h}^{2}a_{h\to l}^2
    \cos\theta
    \nonumber \\
    &+(\mathcal{M}_0^{h(0)})^{2}m_{h}^{2}a_{h\to l}^{2} .
    \nonumber
\end{align}
The scattering length is 
\begin{align}
    |a_{h\to l}|
    = &
    \frac{|B/\cos\theta|}{2m_h\sqrt{A}|\mathcal{M}_0^{h(0)}|} ,
    \nonumber \\
    \frac{a_{h\to l}}{|a_{h\to l}|}
    = &
    \frac{\mathcal{M}_0^{(0)}}{|\mathcal{M}_0^{(0)}|}
    \cdot \frac{\mathcal{M}_0^{h(0)}}{|\mathcal{M}_0^{h(0)}|} 
    \cdot
    \frac{B/\cos\theta}{|B/\cos\theta|} 
    \nonumber .
\end{align}
Thus, the scattering length can be extracted in the same way as before when the relative phase between $\mathcal{M}_{0}$ and $\tilde{\mathcal{M}}_{1}$ is under control. In practice, the relative phase mainly stems from the final-state interactions of the $\pi\pi$ scattering, which are well understood by the phase-shift analysis.

Even if the relative phase is not known in advance, the magnitude of the scattering length can be extracted from the coefficients of the linear terms:
\begin{align}
    |a_{h\to l}|
    = &
    \frac{\sqrt{B^{2}+(B^{\prime})^{2}}}{2m_h\sqrt{A}|\mathcal{M}_0^{h(0)}|}
    \nonumber .
\end{align}
In addition, $\cos\theta$ can be determined from the $\delta^{2}$ terms if we neglect the $\mathcal{M}_0^{(2)}$ term:
\begin{align}
    \cos\theta
    = &
    \frac{\mathcal{M}_0^{h(0)}}{\mathcal{M}_0^{(0)}}
    \frac{C-C^{\prime}}{C+C^{\prime}}
    \quad
    \text{for }
    \mathcal{M}_0^{(2)}= 0
    \nonumber .
\end{align}
In this way, extraction of the scattering length is possible for the complex amplitudes, when information of the final-state interaction and/or high-precision data of the decay spectrum are available.

%%%%%%%%%%%%%%%%%%%%%%%%%%%%%%%%%%%%%%%%%%%%%%%%%%%%%%%%%%%%%%%%%%
\section{Estimation of the mass distribution}\label{sec:estimation}
%%%%%%%%%%%%%%%%%%%%%%%%%%%%%%%%%%%%%%%%%%%%%%%%%%%%%%%%%%%%%%%%%%

So far we have discussed how to extract the $\pi\Sigma$ scattering lengths from the observed decay spectrum. In this section, we present  theoretical estimation of the mass spectrum~\eqref{eq:ampsquare2}, in order to examine the sensitivity of the cusp effect on the $\pi\Sigma$ scattering lengths. 

%------------------------------
\subsection{Input parameters}
\label{subsec:input}

To evaluate the mass spectrum, we need to determine the amplitudes for the weak process and the value of the scattering length. We first consider the weak process of the $\Lambda_c\to \pi\pi\Sigma$. Since we are not aiming at the construction of the realistic decay spectrum, we simply approximate the amplitude by the leading-order constant $\mathcal{M}_0=\mathcal{M}_0^{(0)}$ and $\mathcal{M}_0^h= \mathcal{M}_0^{h(0)}$, which would be sufficient for the present purpose. We first put $\theta = 0$ and the effect of the relative phase will be considered in the end of this section. The magnitude of these amplitudes are determined by the partial decay widths derived from the central values of the lifetime of the $\Lambda_c$ and the branching ratios given in Table~\ref{tbl:ratios}. The results are summarized in Table~\ref{tbl:weak}. For the $a^{0+}$ mode, the branching ratio to the lower energy channel is not known and we assume the same strength with the higher energy channel. In addition, the relative signs of the amplitudes have been determined by evaluating the quark diagrams of the weak decay in Appendix~\ref{sec:weak}. In the following, we adopt the central values for the numerical calculation of the spectra.

\begin{table}[b]
\caption{\label{tbl:weak} Strengths of the weak interaction vertices.}
\begin{ruledtabular}
\begin{tabular}{lll}
mode & $\mathcal{M}_0^{h}$ $(10^{-7}\text{ MeV}^{-1})$
 & $\mathcal{M}_0$ $(10^{-7}\text{ MeV}^{-1})$ \\
\colrule
$a^{-+}$ & $\phantom{-}1.46\pm 0.43$  & $\phantom{-}2.10\pm 0.58$  \\
$a^{00}$ & $\phantom{-}1.46\pm 0.43$  & $-1.49\pm 0.66$ \\
$a^{0+}$ & $-1.49\pm 0.66$ & $-1.49\pm 0.66$\footnote{Assumed to be the same with $\mathcal{M}_0^{h}$} \\
\end{tabular}
\end{ruledtabular}
\end{table}%

Next we consider the $\pi\Sigma$ scattering lengths. It is known that the $\pi N$ scattering length is well described by the leading-order term in ChPT, up to 10\%. Thus, as a first trial, we may adopt the leading-order ChPT to estimate the $\pi \Sigma$ scattering lengths. Higher-order contributions to the scattering lengths are calculated in the heavy baryon formalism~\cite{Liu:2006xja} and in the covariant formalism with the infrared regularization scheme~\cite{Mai:2009ce}. With the definition~\eqref{eq:def}, the leading-order term in ChPT gives the $\pi\Sigma$ scattering length for the channel $h\to l$ as
\begin{align*}
    a_{h\to l}
    =& 
    \frac{M_h}{4\pi(M_h+m_h)}
    \frac{C_{h\to l}(m_h+\omega_l)}{4f^2} , \\
    &\omega_{l}
    =\frac{(M_{h}+m_{h})^{2}-M^{2}_{l}+m^{2}_{l}}{2(M_{h}+m_{h})}
    =m_{l}+\cdots ,
\end{align*}
where $f=92.4$ MeV is the pion decay constant and the ellipsis denotes the isospin breaking correction. The coefficient $C_{h\to l}$ is the group theoretical factor~\cite{Hyodo:2006yk,Hyodo:2006kg}, which takes the values $0$, $2$, and $-2$ for the $a^{-+}$, $a^{00}$, and $a^{0+}$ modes, respectively. Thus, the leading-order ChPT predicts 
\begin{align}
    a^{-+}
    =& 0.00 \text{ fm},\quad
    a^{00}
    =  0.23 \text{ fm},\quad
    a^{0+}
    =  -0.23 \text{ fm}.
    \nonumber
\end{align}

\begin{table}[b]
\caption{\label{tbl:slengths} Estimation of the scattering lengths in units of fm. ChPT stands for the leading-order prediction in chiral perturbation theory. The other results are obtained in the resummation scheme for $I=0$ amplitude when the pole singularities are developed at the energies in parenthesis away from the threshold~\cite{Ikeda:2011dx}.}
\begin{ruledtabular}
\begin{tabular}{lrrrr}
   & $a^{I=0}$  & $a^{-+}$
   & $a^{00}$ & $a^{0+}$ \\
   \colrule
   ChPT           & 0.46    & 0.00    & 0.23    & $-0.23$ \\
   Resonance ($+31$ MeV)     & 1.20    & 0.25    & 0.48    & $-0.23$ \\
   Virtual state ($-6$ MeV) & 5.50    & 1.68    & 1.91    & $-0.23$ \\
   Bound state ($-10$ MeV)   & $-2.30$ & $-0.92$ & $-0.69$ & $-0.23$ \\
\end{tabular}
\end{ruledtabular}
\end{table}%

In the isospin basis, the coupling strengths are given by $4$, $2$, and $-2$ for $I=0$, $1$, and $2$ channels. The strength $C=2$ is the same with the $\pi N(I=1/2)$ channel, so we expect that the ChPT prediction works well. This is indeed the case for the $I=2$ scattering length, since the lattice QCD simulation by NPLQCD Collaboration~\cite{Torok:2009dg} found that $a_{\pi^{+}\Sigma^{+}}=-0.197\pm 0.017$ fm, which is comparable with the leading-order ChPT prediction of $-0.2294$ fm. On the other hand, $C=4$ in the $I=0$ channel is twice as strong as the $\pi N$ interaction so the resummation effect may be important. Indeed, it is found that the chiral $\pi\Sigma(I=0)$ interaction generates a resonance state by the single-channel resummation~\cite{Hyodo:2007jq}. The relation between the pole singularity around the $\pi\Sigma$ threshold and the $\pi\Sigma$ scattering length is studied in Ref.~\cite{Ikeda:2011dx}, which shows that the position and the nature of the pole are sensitive to the value of the scattering length. Depending on the nature of the lower pole of $\Lambda(1405)$, the typical $\pi\Sigma(I=0)$ scattering lengths are found as
\begin{align}
    a^{I=0}
    = 
    \begin{cases}
     1.2 \text{ fm:} & \text{resonance with } E= + 31\text{ MeV} \\
     5.5 \text{ fm:} & \text{virtual state with } E=-6\text{ MeV} \\
     -2.3 \text{ fm:} & \text{bound state with } E=-10\text{ MeV} 
    \end{cases} ,
    \nonumber
\end{align}
where $E$ denotes (the real part of) the pole singularity measured from the threshold.\footnote{The calculation in Ref.~\cite{Ikeda:2011dx} was done in the isospin limit. The resonance case is estimated by taking an average of two different solutions, A1 and B E-dep in Ref.~\cite{Ikeda:2011dx}.} It can be seen that the scattering length takes a large value when the pole is located close to the threshold, and the sign of the scattering length is inverted when a bound state appears. With these values for the $I=0$ scattering length, while the $I=1$ and the $I=2$ components are fixed as the values by the leading-order ChPT, we obtain the scattering lengths by use of Eqs.~\eqref{eq:amp}, \eqref{eq:a00}, and \eqref{eq:a0p} as shown in Table~\ref{tbl:slengths}. Even when $a^{I=0}$ is as large as 5.50 fm, the scattering lengths $a^{-+}$ and $a^{00}$ are not much enhanced, because of the $1/3$ factor in Eqs.~\eqref{eq:amp} and \eqref{eq:a00}. Note that $a^{0+}$ does not depend on the $I=0$ amplitude as observed in its isospin decomposition~\eqref{eq:a0p}.

%--figure---------------------------------
\begin{figure}[tbp]
    \centering
    \includegraphics[height=4.5cm,clip]{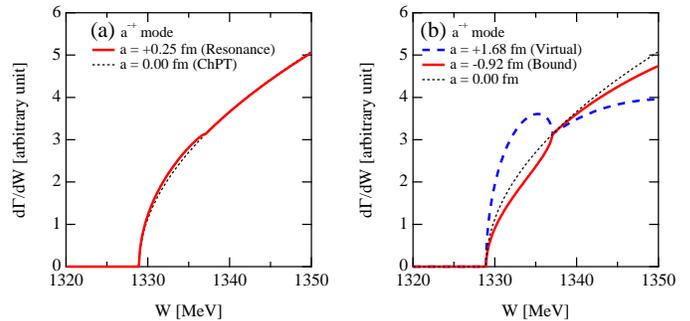}
    \caption{\label{fig:Spectrum1}
    (Color online) Spectra of the $\pi^{-}\Sigma^{+}$ channel in the $\Lambda_{c}\to \pi^{+}(\pi^{-}\Sigma^{+})$ decay with several values of the $a^{-+}$ scattering length.}
\end{figure}%
%--figure---------------------------------

%--figure---------------------------------
\begin{figure*}[btp]
    \centering
    \includegraphics[height=5cm,clip]{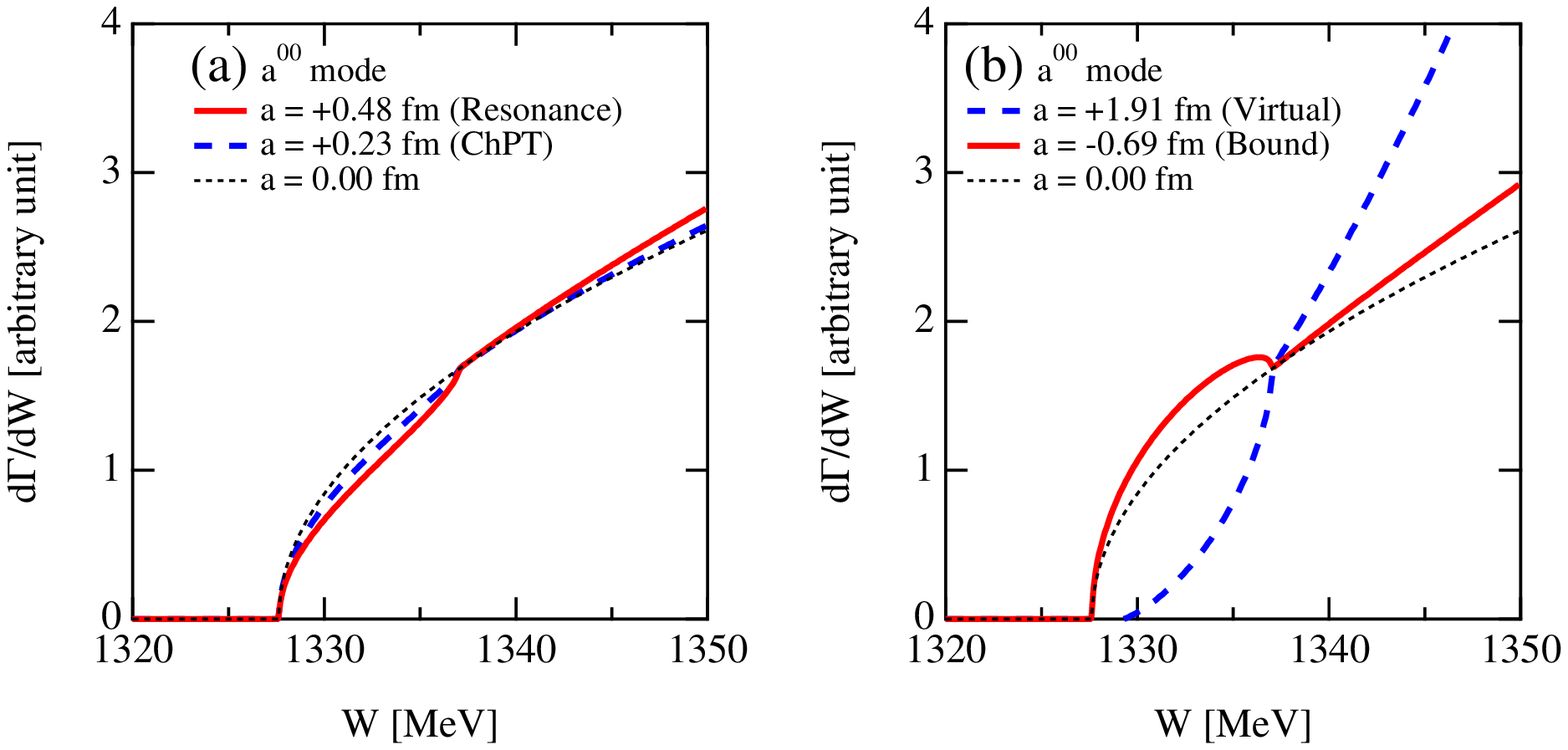}
    \includegraphics[height=5cm,clip]{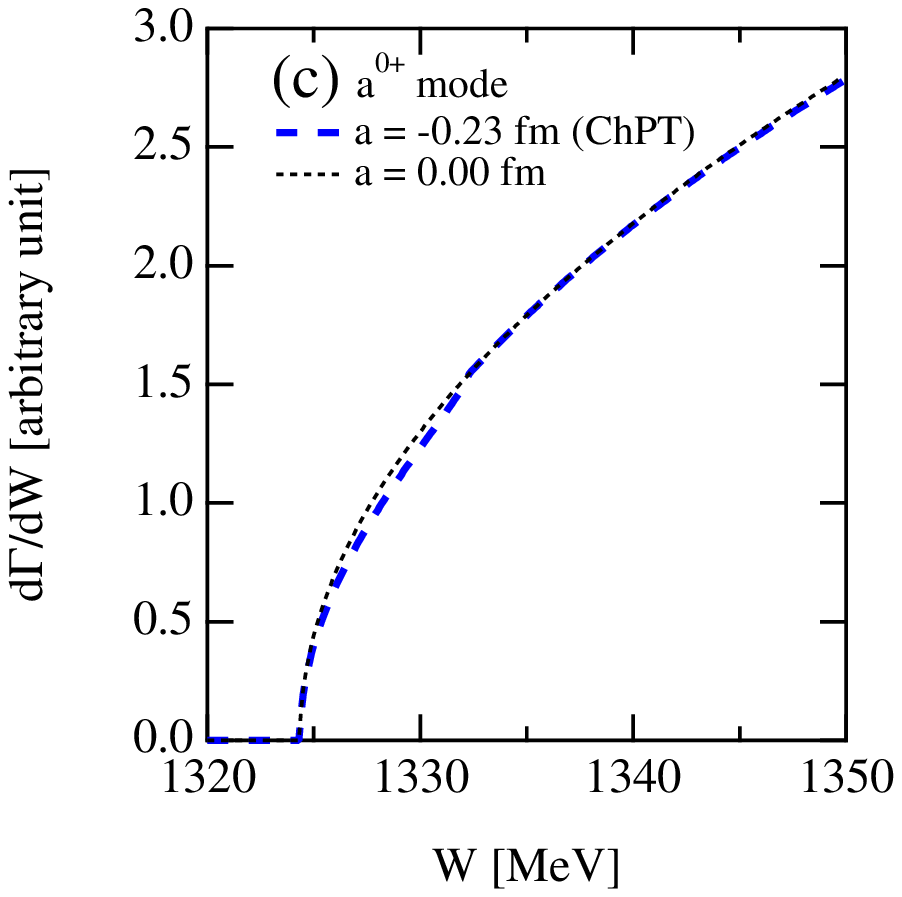}
    \caption{\label{fig:Spectrum2}
    (Color online) Spectra of the $\pi^{0}\Sigma^{0}$ channel in the $\Lambda_{c}\to \pi^{+}(\pi^{0}\Sigma^{0})$ decay  [(a) and (b)] and the $\pi^{0}\Sigma^{+}$ channel in the $\Lambda_{c}\to \pi^{0}(\pi^{0}\Sigma^{+})$ decay (c), with several values of the scattering length. }
\end{figure*}%
%--figure---------------------------------

%------------------------------
\subsection{Examples of $\pi\Sigma$ spectrum}

Using the information given in the previous section, we calculate the $(\pi\Sigma)_{l}$ mass spectrum in the $\Lambda_{c}\to \pi(\pi\Sigma)_{l}$ process by use of Eq.~\eqref{eq:spectrum}. In Fig.~\ref{fig:Spectrum1}, we present the mass spectra of $\pi^{-}\Sigma^{+}$ channel in the $\Lambda_{c}\to \pi^{+}(\pi^{-}\Sigma^{+})$ mode for several values of $a^{-+}$. The dotted lines represent the case when the scattering length is set to be zero. The spectrum shows the pure phase-space distribution and no cusp structure appears, because of the absence of the $B$ term in Eq.~\eqref{eq:ampsquare2}. In this mode, $a^{-+}=0.00$ fm also corresponds to the prediction of the leading-order ChPT, as shown in Table~\ref{tbl:slengths}.

When a finite scattering length is adopted, the cusp structure appears (the solid lines and the dashed line). The cusp effect is not very significant for the case with $a^{-+}=0.25$ fm, but clear structures are visible for $a^{-+}=1.68$ fm and $a^{-+}=-0.92$ fm. In addition, the sign of the scattering length is reflected in the behavior of the lower part of the spectrum in comparison with the phase-space distribution.

Next, in Fig.~\ref{fig:Spectrum2}, we show the spectra of the $\Lambda_{c}\to \pi^{+}(\pi^{0}\Sigma^{0})$ decay and the $\Lambda_{c}\to \pi^{0}(\pi^{0}\Sigma^{+})$ decay with $a^{00}$ and $a^{0+}$, respectively. Again, the dotted lines represent the case when the scattering length is set to be zero. We observe that the prominent cusp structure is seen when the scattering length is as large as 0.5 fm. Because of the difference of the relative sign of the $\mathcal{M}_{0}$ and $\mathcal{M}_{0}^{h}$ amplitudes, the interference pattern in the $a^{00}$ mode is inverted from that in the $a^{-+}$ mode (see Table~\ref{tbl:weak}).

Let us compare three decay modes. In both the $a^{-+}$ and $a^{00}$ modes, the effect of the $I=0$ component is reflected in the cusp structure, so these modes can be used to extract the isosinglet scattering length. A nice feature of the $a^{-+}$ mode is the large branching fraction of the decay ($\sim$3.6\%) which is advantageous for the experimental observation. On the other hand, the scattering length vanishes in the leading-order ChPT, and, hence, the expected scattering length $a^{-+}$ by the resummation effect is not very large. The $a^{00}$ mode has a smaller branching ratio to the final state, but the value of $a^{00}$ is in general expected to be larger than $a^{-+}$ as shown in Table~\ref{tbl:slengths}, which causes a stronger cusp effect. The $a^{0+}$ mode is not useful to extract the $I=0$ component, and the expected cusp structure seems to be relatively weak. In addition, observing two neutral pions in the final state would be experimentally difficult. Nevertheless, once the precise decay spectrum is observed, this mode serves as a precise test of the ChPT prediction for the scattering lengths in $I=1$ and $I=2$ channels. 

\begin{figure}[tbp]
    \centering
    \includegraphics[height=5.5cm,clip]{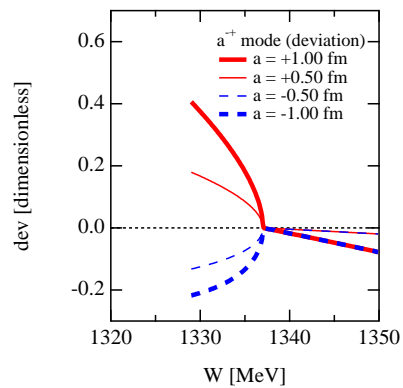}
    \caption{\label{fig:Dev_mp}
    (Color online) Deviations from the phase-space distribution defined in Eq.~\eqref{eq:deviation} for the $\Lambda_{c}\to \pi^{+}(\pi^{-}\Sigma^{+})$ decay with several values of the scattering length $a$.}
\end{figure}

In order to estimate the accuracy of the spectrum measurement  for the feasible determination of the scattering length, we consider the deviation of the spectrum from the phase-space distribution
\begin{align}
    \text{dev}(a)
    = 
    \frac{d\Gamma/dW(a)-d\Gamma/dW(a=0)}{d\Gamma/dW(a=0)}
    \label{eq:deviation} .
\end{align}
For a given $a$, the magnitude of this deviation provides the lower limit of the experimental accuracy to extract the scattering length. We also note that the deviation is always zero at the higher energy threshold $W_{\text{th}}=M_{h}+m_{h}$, so the phase-space distribution can be normalized at this point. In Fig.~\ref{fig:Dev_mp}, we show the deviation plot for the $\Lambda_{c}\to \pi^{+}(\pi^{-}\Sigma^{+})$ decay with several values of the scattering length $a$. We observe that accuracy of about 20\% (10\%) is required to determine the scattering length of $|a|=1$ fm (0.5 fm).

%--figure---------------------------------
%--figure---------------------------------

\begin{figure}[tbp]
    \centering
    \includegraphics[height=4.5cm,clip]{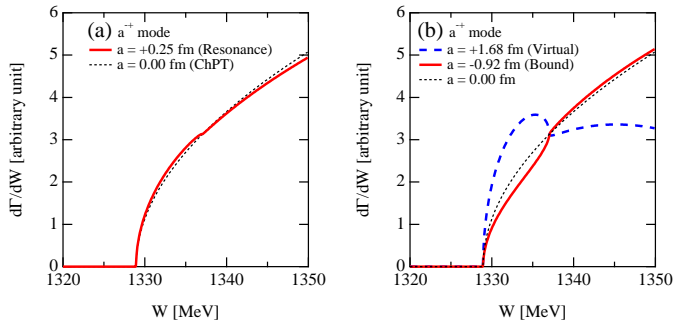}
    \caption{\label{fig:Spectrum_phase}
    (Color online) Spectra of the $\pi^{-}\Sigma^{+}$ channel in the $\Lambda_{c}\to \pi^{+}(\pi^{-}\Sigma^{+})$ decay with several values of the $a^{-+}$ scattering length with the relative phase $\theta=12^{\circ}$.}
\end{figure}

We also study the effect of the relative phase between $\mathcal{M}_{0}$ and $\tilde{\mathcal{M}}_{1}$. As discussed in Appendix~\ref{sec:kinematics}, the main source of the contamination process to the $\Lambda_{c}\to \pi^{+}(\pi^{+}\Sigma^{-})\to \pi^{+}(\pi^{-}\Sigma^{+})$ decay is the $\pi\pi$ rescattering. If the primary $\pi^{+}$ interacts with the $\pi^{+}$  in the intermediate state, the $\tilde{\mathcal{M}}^{1}$ has an imaginary part. This is the $\pi\pi$ scattering in the isospin $I=2$ channel. The relevant energy region of the $\pi\pi$ invariant mass is about 400--800 MeV, and the phase shift of this $\pi\pi$ channel ranges between $-5^{\circ}$ to $-20^{\circ}$, with almost linear energy dependence~\cite{Oller:1998hw}. Although the phase depends on the invariant mass of the $\pi\pi$ pair, we may represent the $\pi\pi$ rescattering effect by a single energy-independent relative phase. Thus, for a rough estimation, we introduce an energy-independent relative phase $\theta=-12^{\circ}$ and calculate the spectrum with the formulas in Sec.~\ref{subsec:extension}. The results are shown in Fig.~\ref{fig:Spectrum_phase}. One observes that the spectrum is slightly modified from Fig.~\ref{fig:Spectrum1}, especially above the threshold. The existence of the relative phase affects the interference pattern of the two amplitude, while the cusp structure itself remains, as discussed in Sec.~\ref{subsec:extension}.

%--figure---------------------------------
%--figure---------------------------------

%%%%%%%%%%%%%%%%%%%%%%%%%%%%%%%%%%%%%%%%%%%%%%%%%%%%%%%%%%%%%%%%%%%%%%%%
\section{Summary}\label{sec:summary}
%%%%%%%%%%%%%%%%%%%%%%%%%%%%%%%%%%%%%%%%%%%%%%%%%%%%%%%%%%%%%%%%%%%%%%%%

We study the $\pi\Sigma$ mass distribution in the $\Lambda_{c}\to \pi\pi\Sigma$ decays to extract the $\pi\Sigma$ scattering lengths. The threshold cusp effect is discussed in detail, and the relation between the $\pi\Sigma$ scattering length and the mass distribution is derived. We show that the $\pi\Sigma$ scattering length can be extracted by the expansion of the observed spectrum around the cusp structure. 

We then evaluate the $\pi\Sigma$ spectrum for several values of the scattering lengths obtained by chiral perturbation theory and the resummation technique. It is found that, if the scattering length is as large as $0.5$ fm, a substantial cusp structure will be observed with about 10 \% deviation in the experimental data. This value of the scattering length corresponds to the case when the $I=0$ $\pi\Sigma$ scattering amplitude has a bound or virtual state about 10 MeV below the threshold. Thus, the analysis of the cusp structure will give important information on the $\pi\Sigma$ interaction around the threshold.

Note, however, that we have presented a rough estimation of the spectrum and there are several other contributions to the same process which are not included in the present calculation. A detailed Dalitz analysis of the $\Lambda_{c}\to \pi\pi\Sigma$ process is highly desired for the calibration of the theoretical model of the weak process as well as for the investigation of the strong final-state interactions in the $\pi\Sigma$ channel. 

In the $\Lambda_{c}$ decay process, it is possible to provide two constraints on the three different isospin component of the $\pi\Sigma$ scattering lengths. To complete the determination of all three isospin components, for instance, we need to determine the $I=2$ component by the lattice QCD simulation~\cite{Fukugita:1994ve,Torok:2009dg,Ikeda:2010zz}.

\begin{acknowledgments}
The authors are grateful to Dr. Yoichi Ikeda and Professor Wolfram Weise for useful discussions.
T.H.\ thanks the support from the Global Center of Excellence Program by MEXT, Japan, through the Nanoscience and Quantum Physics Project of the Tokyo Institute 
of Technology. 
This work was partly supported by the Grant-in-Aid for Scientific Research from 
MEXT and JSPS (Nos.\
  19540275, 
  21840026, 
  and 22105503).
\end{acknowledgments}

\appendix

%%%%%%%%%%%%%%%%%%%%%%%%%%%%%%%%%%%%%%%%%%%%%%%%%%%%%%%%%%%%%%%%%%
\section{KINEMATICS OF THE $\Lambda_{c}\to \pi\pi\Sigma$ PROCESS}
\label{sec:kinematics}
%%%%%%%%%%%%%%%%%%%%%%%%%%%%%%%%%%%%%%%%%%%%%%%%%%%%%%%%%%%%%%%%%%

Here we summarize the kinematics of the $\Lambda_{c}\to \pi\pi_{l}\Sigma_{l}$ process for the assessment of the possible contamination processes. We consider the threshold energy region of the $\pi_{l}\Sigma_{l}$ pair. In this appendix, we adopt the isospin averaged masses for $\Sigma$ and $\pi$: $M_{\Sigma}=1193$ MeV and $m_{\pi}=138$ MeV.

%--figure---------------------------------
\begin{figure}[tbp]
    \centering
    \includegraphics[width=0.35\textwidth,clip]{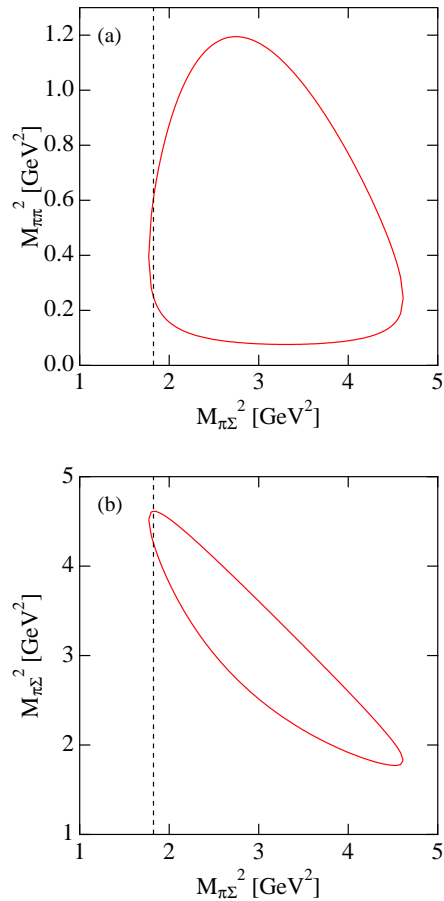}
    \caption{\label{fig:Dalitz}
    Dalitz plots of the $\Lambda_c\to \pi \pi_l \Sigma_{l}$ decay with isospin averaged masses $M_{\Lambda_c}=2286$ MeV, $M_{\Sigma}=1193$ MeV, and $m_{\pi}=138$ MeV. (a) $M_{\pi\pi_l}^2$ vs. $M_{\pi_l\Sigma_{l}}^2$ and (b) $M_{\pi\Sigma_{l}}^2$ vs. $M_{\pi_l\Sigma_{l}}^2$. The vertical dashed lines represent the energy $(M_{\Sigma}+m_{\pi}+20 \text{ MeV})^2$.}
\end{figure}%
%--figure---------------------------------

Let us, first, estimate the typical momenta of the three particles. Consider the extreme case when  the relative momentum of the $\pi_l\Sigma_{l}$ system is zero and the invariant mass of the $\pi_l\Sigma_{l}$ pair is given by $m_{\pi}+M_{\Sigma}$. In this case, the magnitude of the three-momentum of the primary $\pi$ is calculated as
\begin{align}
    |p_{\pi}|
    &
    \sim 747 \text{ MeV}  \nonumber .
\end{align}
This is also the magnitude of the total momentum of the $\pi_l\Sigma_{l}$ pair which goes in the opposite direction to the $\pi$ momentum. With the condition of the vanishing relative momentum, the momenta of $\pi_l$ and $\Sigma_{l}$ is given as
\begin{align}
    |p_{\pi_l}|
    = &
    \frac{m_{\pi}}{M_{\Sigma}+m_{\pi}}|p_{\pi}|
    \sim 77 \text{ MeV} \nonumber , \\
    |p_{\Sigma_{l}}|
    = &
    \frac{M_{\Sigma}}{M_{\Sigma}+m_{\pi}}|p_{\pi}|
    \sim 670 \text{ MeV} \nonumber ,
\end{align}
where a large fraction of the momentum of the $\pi_l\Sigma_{l}$ system is carried by the $\Sigma_{l}$ baryon, because of its heavy mass. 

We next show the Dalitz plots of the $\Lambda_c\to \pi \pi_l \Sigma_{l}$ decay in Fig.~\ref{fig:Dalitz} [Fig.~\ref{fig:Dalitz}(a): $M_{\pi\pi_l}^2$ vs. $M_{\pi_l\Sigma_{l}}^2$; Fig.~\ref{fig:Dalitz}(b): $M_{\pi\Sigma_{l}}^2$ vs. $M_{\pi_l\Sigma_{l}}^2$]. The relevant kinematical region for the $\pi_l\Sigma_{l}$ threshold production is the left edge of the Dalitz plots in these figures. For reference, we put vertical dashed lines at the energy $(M_{\Sigma}+m_{\pi}+20 \text{ MeV})^2$. Figure~\ref{fig:Dalitz}(b) shows that the invariant mass of the ``wrong" $\pi\Sigma_{l}$ pair for this energy region is above 2 GeV, while Fig.~\ref{fig:Dalitz}(a) indicates that the invariant mass of the $\pi\pi_{l}$ pair stays around 600 MeV. This difference is caused by the different momenta of $\pi_{l}$ and $\Sigma_{l}$ discussed above. Since $\Sigma_{l}$ goes with larger momentum than $\pi_{l}$ in the opposite direction to the primary $\pi$, the relative momentum of the $\Sigma_{l}\pi$ pair is larger than that of the $\pi\pi_{l}$ pair. As a consequence, the invariant mass of the $\pi\Sigma_{l}$ system is very large, while that for the $\pi\pi_l$ system is moderate.

\begin{figure}[tbp]
    \centering
    \includegraphics[width=0.45\textwidth,clip]{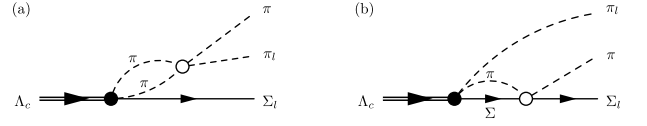}
    \caption{\label{fig:diagramscont}
    Decay diagrams for the $\Lambda_c\to \pi\pi_{l}\Sigma_l$ process: (a) $\pi\pi_{l}$ rescattering and (b) $\pi\Sigma_{l}$ rescattering. Solid circles denote the weak process of $\Lambda_c\to \pi\pi \Sigma$ and the open circles represent the rescattering amplitudes.}
\end{figure}

Based on the kinematical consideration, we discuss the possible contamination processes. In addition to the diagrams in Fig.~\ref{fig:diagrams}, the rescattering of the other combination of the final states may occur, as shown in Fig.~\ref{fig:diagramscont}. In Fig.~\ref{fig:diagramscont}(a) the pions are rescattered, while in Fig.~\ref{fig:diagramscont}(b), the $\Sigma$ is rescattered by the primary pion which is not used to form the mass spectrum. Since the relevant invariant mass of the $\pi\Sigma_{l}$ pair is above 2 GeV where most of the prominent hyperon resonances do not contribute, we may ignore the final-state interaction effect of Fig.~\ref{fig:diagramscont}(b) type. In particular, for the $a^{-+}$ mode, this corresponds to the $I=2$ $\pi^+\Sigma^+$ rescattering, which should be small. On the other hand, the final-state interaction of Fig.~\ref{fig:diagramscont}(a) may be important, since the relevant energy region is near the $\rho$ and $\sigma$ resonances. Because of the isospin, the $\rho$ ($\sigma$) meson can contribute to $a^{-+}$ and $a^{00}$ modes ($a^{-+}$ and $a^{0+}$ modes). The $\sigma$ meson is very broad so the energy dependence of the amplitude is considered to be small. The $\rho$ meson may cause a moderate energy dependence, but the upper limit of the branching ratio of the $\Lambda_c\to \Sigma^+\rho^0$ is given by $<1.4$\%~\cite{\PDG}. Therefore, one may also assume that the effect of the $\rho$ contribution to the present argument is minor. 

We also notice that the diagram in Fig.~\ref{fig:diagramscont} (a) has an imaginary part. If we replace the final state $\pi_{l}\Sigma_{l}$ by $\pi_{h}\Sigma_{h}$, it can be considered as a complex $\tilde{\mathcal{M}}_{1}$ amplitude discussed in Sec.~\ref{subsec:extension}. Since the rescattering takes place by the primary $\pi$ and $\pi_{l}$, the charge combinations are given by
\begin{align}
    a^{-+} :& \pi^+\pi^+ ,
    \quad
    a^{00} : \pi^+\pi^+ ,
    \quad
    a^{0+} : \pi^0\pi^+ .
    \nonumber
\end{align}
In the $a^{0+}$ mode, the rescattering is in $I=1$, which is found to be small as $<1.4$\%~\cite{\PDG}. In the $a^{-+}$ mode and the $a^{00}$ mode, the rescattering is in $I=2$, where the interaction is weakly repulsive. Thus, the effect of the imaginary part is considered to be small for the decay channels of Eqs.~\eqref{eq:amp}, \eqref{eq:a00}, and \eqref{eq:a0p}. This is indeed seen in the numerical estimation in Fig.~\ref{fig:Spectrum_phase}. In any event, to construct a reliable model for the $\Lambda_c\to \pi\pi\Sigma$ decay, detailed analysis of the experimental decay process will be needed.

%--figure---------------------------------
%--figure---------------------------------

%%%%%%%%%%%%%%%%%%%%%%%%%%%%%%%%%%%%%%%%%%%%%%%%%%%%%%%%%%%%%%%%%%%%%%%%
\section{RELATIVE SIGN OF THE WEAK INTERACTION}
\label{sec:weak}
%%%%%%%%%%%%%%%%%%%%%%%%%%%%%%%%%%%%%%%%%%%%%%%%%%%%%%%%%%%%%%%%%%%%%%%%

Here we present the estimation of the relative sign of the weak vertices using the quark diagram method~\cite{Okun:1982ap}. The weak decay of the $c$ quark in $\Lambda_{c}$ occurs through $W^{+}$ exchange as 
\begin{align*}
    c
    \to & s + u + \bar{d},
    \quad
    c+d \to s+u . 
\end{align*}
In addition, it is necessary to create $\bar{q}q$ pair(s) from the vacuum to obtain the $\pi\pi\Sigma$ final state.

Let us first consider the decay processes $\Lambda_{c}\to \pi^{+}(\pi^{-}\Sigma^{+})$ in detail. Four diagrams shown in Fig.~\ref{fig:diag1} can contribute to this process. As shown in Appendix~\ref{sec:kinematics}, the $\pi^{+}$ has a large momentum. This is favored when the $\pi^{+}$ is created by the weak decay of the $c$ quark. In this viewpoint, the contribution from Fig.~\ref{fig:diag1}(d) is considered to be small. In addition, Figs.~\ref{fig:diag1}(b) and \ref{fig:diag1}(c) can be obtained by Fierz rearrangement from Fig.~\ref{fig:diag1}(a) which introduces a suppression factor, as we will see below. 

%--figure---------------------------------
\begin{figure}[tbp]
    \centering
    \includegraphics[width=0.45\textwidth,clip]{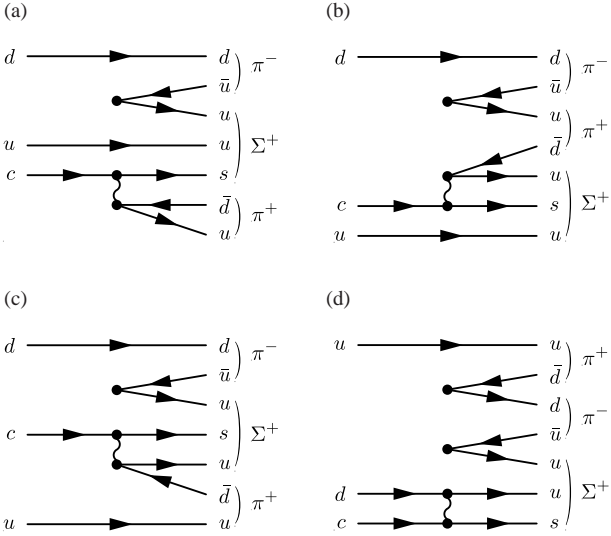}
    \caption{\label{fig:diag1}
    Quark diagrams for the $\Lambda_{c}\to \pi^{+}(\pi^{-}\Sigma^{+})$ decay.}
\end{figure}%
%--figure---------------------------------

The matrix element of the weak process [Fig.~\ref{fig:diag1}(a)] can be written as
\begin{align*}
    \mathcal{M}^{(a)}_{\pi^{+}(\pi^{-}\Sigma^{+})} 
    \sim & 
    \bra{\pi^{+}(\pi^{-}\Sigma^{+})}
    (\bar{u}u) [\bar{u}\Gamma_{\mu}d]
    [\bar{s}\Gamma^{\mu}c]\ket{\Lambda_{c}} ,
\end{align*}
where $\Gamma_{\mu} = \gamma_{\mu}(1-\gamma_{5})$, we insert the operator of the weak interaction, and the $\bar{u}u$ pair is created from the vacuum. The primary pion $\pi^{+}$ has large momentum so it can be factorized as
\begin{align*}
    \mathcal{M}^{(a)}_{\pi^{+}(\pi^{-}\Sigma^{+})}
    \sim & 
    \bra{\pi^{+}}
    \bar{u}\Gamma_{\mu}d
    \ket{0}\bra{\pi^{-}\Sigma^{+}}
    (\bar{u}u)
    [\bar{s}\Gamma^{\mu}c]\ket{\Lambda_{c}}\\
    =& ik_{\mu}^{\pi^{+}}f_{\pi}
    \bra{\pi^{-}\Sigma^{+}}
    (\bar{u}u)
    [\bar{s}\Gamma^{\mu}c]\ket{\Lambda_{c}} ,
\end{align*}
where $k_{\mu}^{\pi^{+}}$ is the momentum of $\pi^{+}$ and $f_{\pi}$ is the pion decay constant. The $\pi^{+}$ is created by the axial vector component of the matrix element. Since the $\pi^{-}$ has a small momentum, we can apply the soft pion theorem to obtain
\begin{align}
    \mathcal{M}^{(a)}_{\pi^{+}\pi^{-}\Sigma^{+}}
    \sim
    &k_{\mu}^{\pi^{+}}
    \bra{\Sigma^{+}}
    \bigl[Q_{5}^{-},(\bar{u}u)\bigr]
    [\bar{s}\Gamma^{\mu}c]\ket{\Lambda_{c}}
    \nonumber  \\
    \sim
    &-k_{\mu}^{\pi^{+}}
    \bra{\Sigma^{+}}
    \bar{u}\gamma_{5}d
    [\bar{s}\Gamma^{\mu}c]\ket{\Lambda_{c}}
    \label{eq:mode1} .
\end{align}

%--figure---------------------------------
\begin{figure}[bp]
    \centering
    \includegraphics[height=2.6cm,clip]{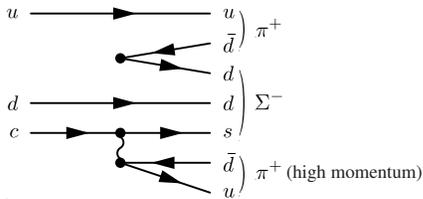}
    \caption{\label{fig:diag2}
    Dominant quark diagram for the $\Lambda_{c}\to \pi^{+}(\pi^{+}\Sigma^{-})$ decay.}
\end{figure}%
%--figure---------------------------------

%--figure---------------------------------
\begin{figure}[tbp]
    \centering
    \includegraphics[width=0.45\textwidth,clip]{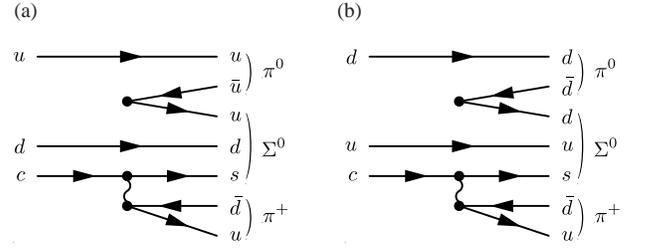}
    \caption{\label{fig:diag3}
    Dominant quark diagrams for the $\Lambda_{c}\to \pi^{+}(\pi^{0}\Sigma^{0})$ decay.}
\end{figure}%
%--figure---------------------------------

Before addressing the other decay channels, we turn to Fig.~\ref{fig:diag1}(b). This amplitude can be obtained by interchanging the $u$ quarks in the amplitude [Fig.~\ref{fig:diag1}(a)]. Noting that the relevant terms to create $\pi^{+}$ are the pseudoscalar and axial vector combinations in the Fierz rearrangement,\footnote{Here we assume that $\bar{u}u$ creation takes place at the same point with the weak interaction vertex, which is not always the case in reality. However, the suppression factor by color and spinor indices as well as the sign of the amplitude are not modified by the nonlocality of the operator.} we obtain
\begin{align*}
    &\bra{\pi^{+}\pi^{-}}(\bar{u}_{1}u)
    (\bar{u}_{2}\Gamma_{\mu}  d_{\beta}) \\
    =&\tfrac{1}{3}
    \bra{\pi^{+}\pi^{-}}
    \Bigl\{
    \tfrac{1}{4}(\bar{u}_{2}\gamma_{\nu}\gamma_{5}d)
    (\bar{u}_{1}\Gamma_{\mu}  
    \gamma^{\nu}\gamma_{5}u) \\
    &-\tfrac{1}{4}(\bar{u}_{2}\gamma_{5}d)
    (\bar{u}_{1}\Gamma_{\mu}  \gamma_{5}u)
    \Bigr\} ,
\end{align*}
where the suppression factor $1/3$ is responsible to obtain the color singlet pair. Thus, the first term is suppressed by the factor 1/12 in comparison with the diagram (a). The second term creates the $\pi^{+}$ from the pseudoscalar operator. The matrix element can be related with that of the axial vector current as
\begin{align*}
    \bra{\pi^{+}}\bar{u}\gamma_{\mu}\gamma_{5}d\ket{0}
    = & 
    \frac{(m_{u}+m_{d})k_{\mu}^{\pi^{+}}}{m_{\pi}^{2}}
    \bra{\pi^{+}}\bar{u}\gamma_{5}d\ket{0} \\
    =& 
    \frac{1}{\chi}
    \bra{\pi}\bar{u}\gamma_{5}d\ket{0} .
\end{align*}
In the case of the hyperon nonleptonic decay, the enhancement factor $\chi$ is large enough to cancel the suppression factor by the Fierz rearrangement~\cite{Okun:1982ap}. In the present case, however, $k_{\mu}^{\pi^{+}}$ is the momentum of the primary pion ($|k_{i}^{\pi^{+}}| \sim 747$ MeV and $k_{0}^{\pi^{+}} \sim 760$ MeV) so the enhancement factor is 
\begin{align*}
    \chi
    =
    \frac{m_{\pi}^{2}}{(m_{u}+m_{d})k^{\pi^{+}}_{\mu}}
    \sim 2.6 ,
\end{align*}
and the pseudoscalar component is not much enhanced. Thus, we consider Fig.~\ref{fig:diag1}(a), in which $\pi^{+}$ is created from the decay of the $W^{+}$ boson, provides the main contribution to this weak process. 

Next, we consider the $\Lambda_{c}\to \pi^{+}(\pi^{+}\Sigma^{-})$ process. The diagram corresponding to Fig.~\ref{fig:diag1}(a) is shown in Fig.~\ref{fig:diag2} where the $\pi^{+}$ generated in the $W^{+}$ decay is regarded as the primary pion. In this case, the $\bar{d}d$ pair should be created from the vacuum instead of $\bar{u}u$. Following the same step as before, we obtain
\begin{align}
    \mathcal{M}_{\pi^{+}(\pi^{+}\Sigma^{-})}
    \sim
    &-k_{\mu}^{\pi^{+}}
    \bra{\Sigma^{-}}
    \bar{d}\gamma_{5}u
    [\bar{s}\Gamma^{\mu}c]\ket{\Lambda_{c}}
    \label{eq:mode2} .
\end{align}
The dominant contributions to the $\Lambda_{c}\to \pi^{+}(\pi^{0}\Sigma^{0})$ process are shown in Fig.~\ref{fig:diag3}. We have two analogous diagrams with $\bar{u}u$ creation [Fig.~\ref{fig:diag3}(a)] and with $\bar{d}d$ creation [Fig.~\ref{fig:diag3}(b)]. From these diagrams, we obtain
\begin{align}
    \mathcal{M}_{\pi^{+}(\pi^{0}\Sigma^{0})}
    \sim
    &-k_{\mu}^{\pi^{+}}
    \bra{\Sigma^{0}}
    (\bar{u}\gamma_{5}u-\bar{d}\gamma_{5}d) 
    [\bar{s}\Gamma^{\mu}c]\ket{\Lambda_{c}}
    \label{eq:mode3} .
\end{align}

%--figure---------------------------------
\begin{figure}[t]
    \centering
    \includegraphics[height=2.6cm,clip]{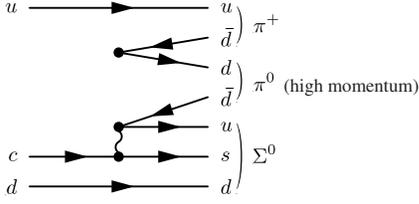}
    \caption{\label{fig:diag4}
    Quark diagram for $\Lambda_{c}\to \pi^{0}(\pi^{+}\Sigma^{0})$ decay.}
\end{figure}%
%--figure---------------------------------

\begin{figure}[tbp]
    \centering
    \includegraphics[height=2.6cm,clip]{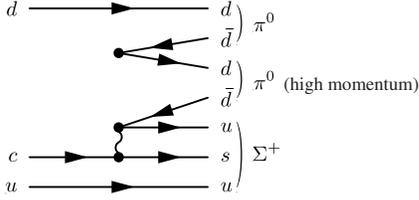}
    \caption{\label{fig:diag5}
    Quark diagram for $\Lambda_{c}\to \pi^{0}(\pi^{0}\Sigma^{+})$ decay.}
\end{figure}

For the processes $\Lambda_{c}\to \pi^{0}(\pi^{+}\Sigma^{0})$ and $\Lambda_{c}\to \pi^{0}(\pi^{0}\Sigma^{+})$, in which the $\pi^{0}$ is the primary pion, there is no corresponding diagram to Figs.~\ref{fig:diag2} and \ref{fig:diag3}. Thus, the leading contribution would be one of the Fierz rearranged diagrams. Here we estimate the relative sign by taking the graphs in Figs.~\ref{fig:diag4} and \ref{fig:diag5} which can be written as
\begin{align*}
    \mathcal{M}_{\pi^{0}(\pi^{+}\Sigma^{0})}
    \sim & \bra{\pi^{0}(\pi^{+}\Sigma^{0})}
    (\bar{d}d)
    [\bar{u}\Gamma_{\mu}d] 
    [\bar{s}\Gamma^{\mu}c]\ket{\Lambda_{c}} , \\
    \mathcal{M}_{\pi^{0}(\pi^{0}\Sigma^{0})}
    \sim & \bra{\pi^{0}(\pi^{0}\Sigma^{+})}
    (\bar{d}d)
    [\bar{u}\Gamma_{\mu}d] 
    [\bar{s}\Gamma^{\mu}c]\ket{\Lambda_{c}} .
\end{align*}
Utilizing the Fierz transformation and soft pion theorem, we obtain
\begin{align}
    \mathcal{M}_{\pi^{0}(\pi^{+}\Sigma^{0})}
    \sim & 
    -\frac{m_{\pi}^{2}}{12(m_{u}+m_{d})}
    \nonumber \\
    &\times \bra{\Sigma^{0}} [\bar{u}\Gamma_{\mu}u
    -\bar{d}\Gamma_{\mu}d ]
    [\bar{s}\Gamma^{\mu}c]\ket{\Lambda_{c}} 
    \label{eq:mode4} , \\
    \mathcal{M}_{\pi^{0}(\pi^{0}\Sigma^{+})}
    \sim & 
    \frac{m_{\pi}^{2}}{12(m_{u}+m_{d})}
    \bra{\Sigma^{+}}
    \bar{u}\Gamma_{\mu}d 
    [\bar{s}\Gamma^{\mu}c]\ket{\Lambda_{c}} 
    \label{eq:mode5} , 
\end{align}
where we have neglected the relatively small axial vector components.

In this way, we obtain the main components of the matrix elements of the weak decay processes. To determine the relative sign, we rotate the matrix elements in the isospin space. The phase convention for the isospin state is given by
%--figure---------------------------------
%--figure---------------------------------
\begin{align}
    \ket{\Sigma^{+}}
    = & 
    -\ket{\Sigma(I=1,I_{3}=1)} 
    \nonumber , \\
    \ket{\pi^{+}}
    = & 
    -\ket{\pi(I=1,I_{3}=1)} 
    \label{eq:phaseconvention} ,
\end{align}
as well as a negative sign for $\bar{d}\gamma_{5}u\sim \pi^{+}$. For instance, it follows from Eq.~\eqref{eq:mode1} that
\begin{align*}
    \mathcal{M}^{(a)}_{\pi^{+}(\pi^{-}\Sigma^{+})}
    \sim
    &-k_{\mu}^{\pi^{+}}
    \bra{\Sigma^{+}}
    \bar{u}\gamma_{5}d
    [\bar{s}\Gamma^{\mu}c]\ket{\Lambda_{c}} \\
    \to
    & 
    -k_{\mu}^{\pi^{+}}
    (-\bra{\Sigma^{-}})
    (-\bar{d}\gamma_{5}u)
    [\bar{s}\Gamma^{\mu}c]\ket{\Lambda_{c}} \\
    =
    & 
    -k_{\mu}^{\pi^{+}}
    \bra{\Sigma^{-}}
    \bar{d}\gamma_{5}u
    [\bar{s}\Gamma^{\mu}c]\ket{\Lambda_{c}} ,
\end{align*}
where the arrow indicates the isospin rotation. Comparing this with Eq.~\eqref{eq:mode2} we find that the amplitude of the $\Lambda_{c}\to \pi^{+}(\pi^{-}\Sigma^{+})$ process has the same sign with that of the $\Lambda_{c}\to \pi^{+}(\pi^{+}\Sigma^{-})$ process. In the same way, from Eq.~\eqref{eq:mode3} we obtain
\begin{align*}
    \mathcal{M}_{\pi^{+}(\pi^{0}\Sigma^{0})}
    \to
    &+k_{\mu}^{\pi^{+}}
    \bra{\Sigma^{-}}
    \bar{d}\gamma_{5}u
    [\bar{s}\Gamma^{\mu}c]\ket{\Lambda_{c}} .
\end{align*}
In comparison with Eq.~\eqref{eq:mode2}, we find the relative sign of $\Lambda_{c}\to \pi^{+}(\pi^{-}\Sigma^{+})$ and $\Lambda_{c}\to \pi^{0}(\pi^{0}\Sigma^{+})$ is odd. Finally, from Eq.~\eqref{eq:mode4} we obtain
\begin{align*}
    \mathcal{M}_{\pi^{0}(\pi^{+}\Sigma^{0})}
    \to &
    +\frac{m_{\pi}^{2}}{12(m_{u}+m_{d})}
    \bra{\Sigma^{+}}
    [\bar{u}\Gamma_{\mu}d ]
    [\bar{s}\Gamma^{\mu}c]\ket{\Lambda_{c}}
    .
\end{align*}
When we compare it with Eq.~\eqref{eq:mode5}, we find the same sign of the processes $\Lambda_{c}\to \pi^{0}(\pi^{+}\Sigma^{0})$ and $\Lambda_{c}\to \pi^{0}(\pi^{0}\Sigma^{+})$. In this way, we determine the relative sign of the weak interaction vertices in Table~\ref{tbl:weak}.

\end{document}